\documentclass[11pt, a4paper]{article}

\usepackage[T1]{fontenc}
\usepackage[utf8]{inputenc}
\usepackage{lmodern}
\usepackage[english]{babel}

\usepackage[a4paper, top=2.5cm, bottom=2.5cm, left=2cm, right=2cm]{geometry}
\usepackage{enumitem}
\setlist[itemize]{label=-}

\usepackage{authblk}
\usepackage{hyperref}
\usepackage{listings}
\usepackage{xcolor}

\definecolor{codegray}{rgb}{0.5,0.5,0.5}
\definecolor{backcolour}{rgb}{0.95,0.95,0.92}

\title{\textbf{PARCER as an Operational Contract to Reduce Variance, Cost, and Risk in LLM Systems}}
\author{Elzo Brito dos Santos Filho\\elzo.filho@cps.sp.gov.br}
\date{February 28, 2026}

\begin{document}

\maketitle

\begin{abstract}
Systems based on Large Language Models (LLMs) have become formidable tools for automating research and software production. However, their governance remains a challenge when technical requirements demand absolute consistency, auditability, and predictable control over cost and latency. Recent literature highlights two phenomena that aggravate this scenario: the stochastic variance inherent in the model's judgment (often treated as "systemic noise") and the substantial degradation of context utilization in long inputs, with critical losses when decisive information is diluted in the middle of the prompt. This article proposes PARCER as an engineering response to these limitations. The framework acts as a declarative "operational contract" in YAML, transforming unstructured interactions into versioned and executable artifacts. PARCER imposes strict governance structured into seven operational phases, introducing decision hygiene practices inspired by legal judgments to mitigate noise, adaptive token budgeting, formalized recovery routes (fallbacks) for context preservation, and systemic observability via OpenTelemetry. The objective of this work is to present the conceptual and technical architecture of PARCER, positioning it as a necessary transition from simple "prompt engineering" to "context engineering with governable governance".
\end{abstract}

\vspace{0.5cm}
\vspace{1cm}

\section{Introduction}

The first wave of technological adoption of Large Language Models (LLMs) was predominantly driven by monolithic \textit{prompting}, where direct textual instructions are provided with the expectation of a coherent final output. While effective for prototyping and exploratory research, this approach reveals profound structural weaknesses when deployed in production environments. In such settings, particularly in digital education and mission-critical software engineering, the narrative plausibility of an AI-generated response is insufficient; procedural governance, cost predictability, and algorithmic traceability are mandatory.

In this production scenario, two operational vulnerabilities stand out. The first is judgment variance. Stochastic systems frequently yield divergent results and asymmetrical arguments for semantically equivalent inputs. The decision psychology literature formalizes this unwanted variability as "noise", arguing that the quality of a corporate or technical judgment must be measured not only by the absence of systematic biases but also by the stability of repeated evaluations. The second vulnerability is the loss of information in inferences involving extensive contexts. Widely documented phenomena, such as "Lost in the Middle" \cite{liu2024lost}, demonstrate that an LLM's capacity to retrieve accurate data drops drastically when crucial information is not located at the edges (beginning or end) of the provided document, thereby invalidating the premise that simply expanding the context window solves knowledge injection issues.

This article proposes PARCER (Prompt Architecture for Cognitive Engineering \& Responses) as a framework for risk containment and reliability enhancement. PARCER functions not as a linguistic refinement of \textit{prompts}, but as a declarative operational contract in YAML. It structures the execution of complex tasks by enforcing dynamic budget limits, escape routes against hallucinations caused by context size, and a rigorous "decision hygiene" that requires evidence substantiation prior to the approval of any generated artifact.

\section{Theoretical Foundation: From Prompting to Controlled Process}

The shift from a "free generation" regime to one of "governed engineering" is supported by multidisciplinary concepts bridging decision psychology, computational efficiency, and software architecture.

\subsection{Decision Hygiene and Systemic Noise}
The concept of "decision hygiene," popularized by Daniel Kahneman, Olivier Sibony, and Cass Sunstein \cite{kahneman2021}, suggests that noise reduction in complex human judgments is achieved not through intuition, but via the application of structured processes: task decomposition, adoption of closed rubrics, and blind aggregations. Transposing this principle to Generative AI, the technical parallel requires decoupling the text generation stage from the rigorous validation stage. Contemporary research on LLM application in the legal sector, such as the \textit{Soppia} framework \cite{araujo2025}, demonstrates that applying direct and inverse logic to heterogeneous criteria forces the model to execute traceable proportional weightings, deterring generic responses. PARCER absorbs this theory, establishing that AI should not merely deliver a result, but rather construct and audit the rationale supporting it.

\subsection{The Fallacy of Unlimited Context}
The promise of LLMs featuring massive context windows created the illusion that the complexity of information retrieval had been resolved. However, empirical studies reveal that the model's attention distribution forms a "U" curve, prioritizing the top and bottom of the text while ignoring facts buried in the middle. In practice, robust enterprise architectures cannot simply "paste" entire databases into the \textit{prompt}. A defensive hierarchy is required, dynamically applying Map-Reduce summaries and Retrieval-Augmented Generation (RAG) to compress information density before the degradation threshold of the model's attentional memory is surpassed.

\subsection{Inference Efficiency and Dynamic Budgets}
Another latent challenge is computational \textit{overthinking}: LLMs frequently expend excessive tokens and time on trivial tasks when embedded in agentic loops. Recent literature concerning dynamic resource allocation (e.g., studies surrounding DiffAdapt \cite{diffadapt} and ABPO \cite{abpo}) proposes that a model's reasoning budget should be flexible, scaling proportionally with the statistical uncertainty of the prediction or the perceived complexity of the task. This adaptability inspired PARCER's real-time budgeting module.

\section{The Operational Architecture of PARCER}

PARCER materializes governance by requiring that all LLM behavior be bounded by a structured YAML artifact. Rather than relying on the model to "guess" the scope and limits of its operation, the contract explicitly specifies the engine, reasoning phases, budgets, and safety valves. The operational flow and internal mechanisms of the framework are detailed below.

\subsection{Base Configuration: Metadata, Preambles, and Conversational State}
The operation begins with the initialization of the \textit{prefaces} block, which binds telemetry, fixed maximum limits (such as latency in milliseconds and maximum token budget), and guardrails. A vital feature of this version is the memory reuse strategy via the \textit{Responses API} \cite{openai-responses}. Instead of the traditional and costly transmission of the entire conversational history at each new iteration, PARCER utilizes previous response identifiers (\textit{previous\_response\_id}). This logical chaining not only conserves bandwidth and lowers operational costs but also establishes deterministic traceability, ensuring the agent maintains its context state without polluting the input window with redundant metadata.

\subsection{The Seven Phases of Strict Execution}
To prevent reasoning collapse, PARCER divides the workflow into a non-negotiable taxonomy of seven phases. In Phase 1 (Analysis), the model is compelled to rewrite the user's objective and select an engagement mode (e.g., "high persistence" versus "low anxiety" context gathering). Phase 2 (Plan) requires mapping operational risks before triggering any tools. Phase 3 (Execution) permits external actions under strict formatting rules and batch iteration limits. Phase 4 (Validation) is addressed separately in the following subsection, given its structural significance. Following judgment, Phase 5 (Review) triggers the generation of structured metrics. Finally, Phases 6 and 7 format the delivery (Handoff) complete with an audit \textit{changelog} and a summary of assumed hypotheses.

\subsection{Phase 4: Decision Hygiene as the Heart of the Framework}
The most profound innovation of PARCER lies in Phase 4. Diverging from conventional "review your work" instructions, Phase 4 establishes an internal tribunal. The model must compile an \textit{Evidence Docket} (linking key assertions to retrieved sources) and an \textit{Assumptions Register} (an explicit log of hypotheses where empirical evidence was lacking).

The framework applies seven Decision Hygiene procedures (DH1 to DH7) to these artifacts. These procedures dictate that mathematical quality evaluation must be non-linear. Inspired by legal frameworks for damage valuation \cite{araujo2025}, PARCER employs inverse logic for the efficiency criterion: the more tokens, latency, and cost required to complete the task, the lower the assigned score. Beyond weighted scoring (where Factual Fidelity and Fitness carry the most weight), the system enforces \textit{Hard Gates}. If the score for the "Safety" or "Factual Fidelity" gate fails to meet a non-negotiable threshold, the contract prohibits delivery of the result, triggering a forced replanning routine or returning a rejection verdict. Additionally, counterfactual tests and adversarial probes are incorporated, compelling the LLM to actively attempt to invalidate its own conclusion before approving it.

\subsection{Adaptive Budgeting and Fallback Nets}
To prevent the rigor of Phase 4 from rendering the framework prohibitively expensive, the \textit{Adaptive Budgeting} module operates concurrently. Based on a Model Uncertainty Score (MUS), the system calculates whether the current task requires more leeway or should be curtailed. Using exponential moving average (EMA) parameters alongside heating/cooling thresholds (\textit{mus\_heat} and \textit{mus\_cool}), the agent can temporarily expand its "tool calls" limit if investigating a critical issue, or halt execution immediately in economy mode if uncertainty remains high, signaling that the AI is operating blindly.

Accompanying this dynamic budgeting are context \textit{Fallbacks}. PARCER features rigid trigger limits (e.g., contexts exceeding 120,000 characters). Upon crossing this threshold, or if the semantic similarity metric of the search begins to degrade, the system refrains from injecting raw content into the prompt. Instead, it mandates a defensive cascade: first, a Map-Reduce flow for synthetic summarization; if insufficient, it activates a hyper-focused RAG mechanism via Maximal Marginal Relevance (MMR) to extract only high-value excerpts; as a final resort, it forces an early termination, declaring methodological limitations ("partial completion").

\subsection{Central Observability and Multi-Agent Orchestration}
In enterprise production systems, an LLM operating blindly without systemic monitoring presents an unsustainable risk. PARCER resolves this by adopting native semantic tracing compatible with the OpenTelemetry (OTel) standard \cite{otel-genai}. By linking \textit{trace\_ids} to correlation identifiers, each of the seven operational phases becomes a measurable \textit{span} within modern APM systems. This enables detailed exporting of gate failures, dynamic budget adjustments, and precise costs per execution block. Furthermore, the YAML includes explicit integration contracts (\textit{node\_specs} and \textit{edge\_specs}) for graphical orchestration frameworks, enabling PARCER to serve as the foundational regulation for "nodes" in systems built with LangGraph or CrewAI.

\section{Applications: Education and Software Engineering}

While agnostic to specific knowledge domains, the governance capabilities of PARCER exhibit unparalleled value in high-informational-impact contexts. 

In \textbf{Education}, the risk of unrestricted LLM usage manifests as factual hallucinations that distort learning pathways or the generation of assessments failing to reflect original criteria. Through its internal \textit{education profile}, PARCER dynamically adjusts its validation mesh, integrating rigorous Accessibility (A11y) criteria as mandatory approval gates. Furthermore, by compiling the assumptions register and the textual evidence docket, the model provides educational institutions with absolute traceability regarding why a specific grade or didactic suggestion was generated for a student, thereby enabling transparent and consistent pedagogical audits.

In \textbf{Software Engineering}, the code editing rules embedded within the framework prohibit "code golf" (obscure optimizations) and enforce strict design system and componentization premises. The methodological \textit{swap} in quality review changes the classic traceability metric to "dx\_maintainability" (developer experience and maintainability), ensuring that diffs generated by the agent not only compile but also maintain a clean taxonomy, include adequate tests, and do not exacerbate the technical debt of the target codebase.

\section{Conclusion}

The productive adoption of Generative Artificial Intelligence has moved beyond casual interactions based on loose descriptive texts. For language models to deliver safe and sustainable value, they must be managed as rigorous computational processes rather than mere probabilistic oracles. PARCER concretizes this vision by proposing a readable, deterministic, and adaptive machine contract. By demanding auditable decision hygiene to verify the quality of complex responses, autonomously guarding against attentional degradation, and tracking execution via industry-standard telemetry, the framework offers a pragmatic horizon. The objective is not to stifle the algorithmic model's creativity, but to frame it within boundaries that allow context engineering to mature into transparent, sustainable, and auditable corporate governance.

\newpage
\appendix
\section{Appendix: The Full PARCER Operational Contract}

\begin{lstlisting}[language=bash]
PARCER_v1_4_7:
  version: "1.4.7"
  rationale: >
    Consolidates simplified layers (mini-PARCER), adaptive budget control
    (Adaptive Budgeting), context fallbacks (Summaries/RAG), rubric with aggregated
    quality export (review_metrics), connectors for CrewAI/LangGraph with
    central observability, and decision hygiene in Phase 4. Maintains the 1.4.6 backbone
    (agentic, guardrails, memory reuse) with enhanced governance.

  meta:
    title: "<task title>"
    owner: "<owner>"
    date: "<YYYY-MM-DD>"
    tags: [agentic, responses_api, gpt-5, coding, research, observability]

  prefaces:
    model_profile:
      model_family: "gpt-5"
      reasoning_effort: "medium"          # low|medium|high
      verbosity:
        default: "low"                    # concise by default
        code_blocks: "high"               # diffs and detailed technical snippets
      eagerness_mode: "low"               # default; can be switched by switches.agentic_mode
      tool_call_budget:
        max_calls: 6
        parallel_batches: 1
      determinism:
        seed: null
        temperature: 0.2                  # consistency; increase for exploration
    telemetry:
      capture: true
      fields: [correlation_id, response_id, response_time_ms, response_tokens, response_cost]
      cost_budget_tokens: 2000
      latency_budget_ms: 60000
      price_per_token_est: "<USD/token_est>"
    memory_reuse:
      strategy: "responses_api"           # off|summaries|RAG|responses_api
      previous_response_id: "{{previous_response_id}}"
      traceability: "log:correlation_id; changelog at the end"
    guardrails:
      stop_conditions:
        - "If response_cost > cost_budget_tokens * price_per_token_est => stop and report overrun."
        - "If response_time_ms > latency_budget_ms and task is not critical => return partial + next steps."
        - "Destructive/irreversible actions require explicit confirmation."
      uncertainty_thresholds:
        default: "medium"
        tools:
          delete_file: "low"
          pay/checkout: "low"
          search: "high"
          read_only: "high"
      safety_policy:
        - "Do not log sensitive PII."
        - "Cite sources when using the web."
    agentic_preambles:
      - name: "tool_preambles_default"
        content: |-
          <tool_preambles>
          - Rephrase the goal (1-2 lines) before any tool call.
          - List a short, numbered plan.
          - Narrate progress in ticks: [1/3], [2/3],... without excessive verbosity.
          - Close with "Summary of what was done" + assumptions/limitations.
          </tool_preambles>
    agentic_modes:
      low_eagerness: |-
        <context_gathering>
        Goal: Obtain sufficient context quickly and act.
        Method: Max 1 parallel batch; deduplicate searches; stop early (~70% convergence).
        Budget: Absolute maximum of 2 tool calls.
        Fallback: If uncertain, do ONE refined batch and proceed; document assumptions.
        </context_gathering>
      high_eagerness: |-
        <persistence>
        You are an agent: persist until resolved. Confirm only for irreversible actions.
        </persistence>

  scope:
    topic_domain: "<domain/theme>"
    success_criteria:
      - "<criterion_1>"
      - "<criterion_2>"
    constraints:
      - "<constraint_1>"
      - "<constraint_2>"
    assumptions:
      - "<assumption_1>"
      - "<assumption_2>"

  # ---------------- Operational Core: Phases ----------------
  phases:
    "1_analysis":
      steps:
        - "Rewrite user goal (1 line)."
        - "Select eagerness_mode (low|high) and justify in 1 line."
        - "Declare minimum necessary info + tool budget."
    "2_plan":
      steps:
        - "Short plan (3-6 steps), including validation stage and possible fallbacks."
        - "Map risks/uncertainties and how to mitigate them."
    "3_execution":
      rules:
        - "Respect tool budget; avoid repeated searches."
        - "Apply agentic_preambles.tool_preambles_default."
      code_editing_rules: |-
        <code_editing_rules>
        <guiding_principles>
        - Clarity & Maintainability; descriptive names; simple flow; no code-golf.
        - Consistency with existing design system.
        - Componentization for repeated patterns.
        - Diffs ready for review with succinct explanation of 'why'.
        </guiding_principles>
        <frontend_stack_defaults>
        - Next.js (TypeScript), Tailwind, shadcn/ui, Lucide, Zustand
        </frontend_stack_defaults>
        <ui_ux_best_practices>
        - 4-5 sizes/weights; 1 neutral + up to 2 accents; spacing x4; skeletons; a11y Radix/shadcn.
        </ui_ux_best_practices>
        <diff_quality>
        - Explain intentions/contracts; avoid 1-letter variables; comment ambiguous points.
        </diff_quality>
        </code_editing_rules>
      zero_to_one_booster: |-
        <self_reflection>
        - Internally create a rubric with 5-7 criteria; iterate to max score before emitting.
        </self_reflection>
    "4_validation":    # Decision Hygiene (inspired by Soppia 2.1)
      artifacts:
        evidence_docket:
          fields: [evidence_id, source, retrieved_at, excerpt, relevance_note]
        assumptions_register:
          fields: [assumption_id, statement, impact_low_med_high, mitigation]
        derivation_trace:
          fields: [objective_item, supporting_evidence_ids]
      decision_hygiene:
        procedures:
          - id: DH1_evidence_alignment
            goal: "Check factual adherence (faithfulness)."
            checks:
              - "Key assertions mapped to evidence_id?"
              - "Dates/units/names consistent with sources?"
          - id: DH2_criteria_scoring
            goal: "Score criteria 0-5 (direct/inverse according to criterion)."
          - id: DH3_counterfactuals
            goal: "Test 1-2 minimum scenarios that would invalidate the conclusion."
          - id: DH4_adversarial_probes
            goal: "Generate 3 adversarial questions and succinct answers."
          - id: DH5_uncertainty_register
            goal: "Classify uncertainties (epistemic/aleatoric) and impact."
          - id: DH6_variance_check
            goal: "Calculate deviation (Noise Index) in double evaluation, if any."
          - id: DH7_signoff_gates
            goal: "Apply gates and decide Accept/Revise/Reject."
        scoring_model:
          criteria:
            - {id: fitness,       logic: direct}
            - {id: faithfulness,  logic: direct}
            - {id: completeness,  logic: direct}
            - {id: safety,        logic: direct}
            - {id: efficiency,    logic: inverse}   # higher cost/latency => lower score
            - {id: clarity,       logic: direct}
            - {id: traceability,  logic: direct}
          weights: {fitness:0.25, faithfulness:0.20, completeness:0.15, clarity:0.10, efficiency:0.10, safety:0.10, traceability:0.10}
          gates:
            - {id: faithfulness, min_score: 3}
            - {id: safety,       min_score: 3}
            - {id: fitness,      min_score: 3}
        metrics_export:
          fields:
            - scores                # map criterion->0..5
            - weighted_score        # 0..100
            - gates_status          # pass/fail per gate
            - verdict               # accept|revise|reject
            - uncertainty_summary   # <= 3 items
            - actions_required      # prescriptive bullets for rework
            - telemetry_brief       # tokens,cost,latency,tool_calls
            - evidence_refs         # ids/links/hashes
      profiles:
        research:
          extras:
            - "Mandatory evidence_docket with dates and DOIs/URLs."
            - "Additional gate: traceability >= 3."
        coding:
          replacements:
            - "traceability -> dx_maintainability"
          gates_extra:
            - {id: tests_or_sanity, min_score: 3}
        education:
          extras:
            - "Add Accessibility/A11y (0-5) with 0.10 weight (redistribute)."
    "5_review":
      metrics:
        correlation_id: "{{correlation_id}}"
        response_id: "{{response_id}}"
        response_time_ms: "{{response_time_ms}}"
        response_tokens: "{{response_tokens}}"
        response_cost: "{{response_cost}}"
      review_rubric:
        catalog_id: "parcer-core-v1"
        export: true            # enables aggregated export (no chain of thought)
      adaptive_actions:
        - when: "response_tokens > telemetry.cost_budget_tokens"
          do: "lower_verbosity_and_retry"
        - when: "response_time_ms > telemetry.latency_budget_ms"
          do: "skip_non_critical_tools"
        - when: "tool_calls > prefaces.model_profile.tool_call_budget.max_calls"
          do: "collapse_search_paths_and_finalize"
    "6_handoff":
      deliverables:
        - "Summary of what was done."
        - "Assumptions and limitations."
        - "Next steps (optional)."
        - "Diffs/patches or application instructions."
        - "review_rubric.metrics + telemetry_brief (if export active)."
    "7_changelog":
      format: "- {{timestamp}} -- {{change}}"

  # ---------------- Adaptive Budgeting (dynamic control) ----------------
  adaptive_budgeting:
    enabled: true
    targets:
      latency_ms: 60000
      cost_usd: 0.50
      min_quality: 0.75
    caps:
      max_tokens_cap: 2400
      max_tools_cap: 6
    gains:
      tokens_up_pct: 0.15
      tokens_down_pct: 0.20
      tool_step: 1
    thresholds:
      mus_heat: 30           # heat up (increase budget)
      mus_cool: 10           # cool down (reduce budget)
      guard_soft_pct: 0.80   # economy mode
      guard_hard_pct: 1.00   # safe stop
    smoothing:
      ema_alpha: 0.4
      hysteresis_pct: 0.10
      cooldown_steps: 1
    knobs:
      allow_use_web_drop: true
      allow_reasoning_downgrade: true

  # ---------------- Fallbacks (Summaries/RAG/Partial) ----------------
  fallbacks:
    enabled: true
    triggers:
      economy_mode: true              # triggered by Adaptive
      max_context_chars: 120000
      min_top1_sim: 0.35
      min_avg_topk_sim: 0.30
      topk: 5
    actions:
      order: ["summary", "rag", "rag_then_summary", "partial"]
      summary:
        target_tokens: 800
        strategy: "map_reduce"
        map_chunk_chars: 8000
        reduce_passes: 1
      rag:
        k: 5
        mmr_lambda: 0.3
        min_sim: 0.30
        cache_dir: "./ragcache"       # embeddings.npy / meta.json
      partial:
        stop_when_confidence: 0.85

  # ---------------- Connectors & Observability ----------------
  connectors:
    langgraph:
      enabled: false
      graph_contract:
        node_spec_fields: [id, kind, action_contract, budget, policies]
        edge_spec_fields: [from, to, condition]      # quality, gates, budget_remaining, validator_pass
      runtime_policies:
        max_steps: 50
        retry: 1
        backoff: "exp"
        parallel_branches: 2
      observability:
        hooks: [on_node_start, on_node_end, on_edge_transition, on_error]
    crewai:
      enabled: false
      roles:      # [{id, purpose, capabilities, limits, review_expectations}]
      task_contract:
        acceptance_criteria: "mirror rubric"
        handoff_contract: "required artifacts"
        fallback_rights: "summary/rag when economy_mode"
      supervision:
        reviewer_role: "reviewer"
        escalation_rules: "gate failure => replan"
        max_rounds: 3
  observability_central:
    tracing:
      standard: "otel"
      trace_id: "{{correlation_id}}"
      span_format: "phase|action"
      events_enabled: true
    metrics_export:
      include: [review_rubric, telemetry_brief, failures, budget_adjust_log]
    privacy:
      pii_redaction: "on"
      sampling_rate: 0.5
    retention:
      run_logs_ttl_days: 30
      review_exports_ttl_days: 120

  # ---------------- Switches, Profiles, IO ----------------
  switches:
    agentic_mode: "low"                   # low|high
    use_web_search: true
    allow_partial_completion: true
    produce_citations: "when_web_used"
    produce_review_export: true
  profiles:
    research:
      default_mode: "thorough"
      use_web: true
      extra_gates: [{id: traceability, min_score: 3}]
    coding:
      default_mode: "quick"
      use_web: false
      review_swap: "traceability->dx_maintainability"
    education:
      default_mode: "thorough"
      use_web: true
      add_criterion: {id: a11y, weight: 0.10}
  inputs:
    context_input: "<problem description/artifacts>"
    artifacts:
      - name: "<file_or_link>"
        type: "<type>"
  outputs:
    structured:
      type: "json|yaml|diff|md"
      schema_hint: "<optional: target schema>"
    logging:
      include: [plan, tool_calls_summary, validation_results]

  # ---------------- Useful Snippets ----------------
  snippets:
    responses_api:
      example: |-
        {
          "model": "gpt-5",
          "reasoning_effort": "medium",
          "previous_response_id": "{{previous_response_id}}"
        }
    low_latency_profile:
      temperature: 0.0
      reasoning_effort: "low"
      tool_call_budget:
        max_calls: 2
    mini_parcer_embedded: |-
      PARCER_mini_v0_1:
        version: "0.1"
        purpose: "Orchestrate tasks with LLM simply and auditably."
        modes:
          quick: {reasoning_effort: low, verbosity: low, tool_call_budget: {max_calls: 2}, stop_when_confidence: 0.8}
          thorough: {reasoning_effort: medium, verbosity: medium, tool_call_budget: {max_calls: 4}, stop_when_confidence: 0.9}
        guardrails:
          - "No irreversible actions without confirmation."
          - "Avoid sensitive PII."
          - "Cite sources when using web."
          - "Deduplicate searches."
        inputs:
          meta: {title: "<title>", owner: "<owner>", date: "<YYYY-MM-DD>", tags: []}
          context: "<essential facts>"
          goal: "<deliverable>"
          constraints: []
          preferences: {mode: quick, use_web: false, max_tokens: 1200}
        phases:
          analysis: {steps: ["Rewrite goal (1 line).", "Min info missing.", "Declare budget."]}
          plan: {steps: ["3-5 step plan.", "2-3 risks + mitigation."]}
          execution:
            rules:
              - "If using tools, state the goal of each call in 1 line."
              - "Consolidate results and stop at stop_when_confidence."
          validation:
            checks: ["Goal coverage.", "Logical sanity.", "Citations if use_web=true.", "Uncertainties (<=3)."]
          handoff:
            deliverables: ["Summary (4-6 lines).", "Limitations.", "Next steps (optional)."]
        outputs:
          structured: {type: md, include: [analysis, plan, result, validation, handoff]}
        profiles:
          research: {default_mode: thorough, use_web: true}
          coding: {default_mode: quick, use_web: false}
\end{lstlisting}


\vspace{1cm}
\begin{thebibliography}{99}

\bibitem{kahneman2021}
Kahneman, D., Sibony, O., \& Sunstein, C. R. (2021). \textit{Noise: A Flaw in Human Judgment}. Little, Brown Spark.

\bibitem{liu2024lost}
Liu, N. F., Lin, K., Hewitt, J., Paranjape, A., Bevilacqua, M., Petroni, F., \& Liang, P. (2024). \textit{Lost in the Middle: How Language Models Use Long Contexts}. Transactions of the Association for Computational Linguistics (TACL), 12:157--173.

\bibitem{araujo2025}
Araujo, J. A. (2025). \textit{Soppia: A Structured Prompting Framework for the Proportional Assessment of Non-Pecuniary Damages in Personal Injury Cases}. arXiv preprint arXiv:2510.21082v1.

\bibitem{diffadapt}
Liu, X., Hu, X., Chu, X., \& Choi, E. (2026). \textit{DiffAdapt: Difficulty-Adaptive Reasoning for Token-Efficient LLM Inference}. In Proc. of ICLR 2026.

\bibitem{abpo}
Lin, W., Wu, Q., Yang, S., Zhou, Y., Sun, X., \& Ji, R. (2025). \textit{Towards Efficient Chain-of-Thought Reasoning via Adaptive-Budgeting based Policy Optimization}. ICLR 2026.

\bibitem{openai-responses}
OpenAI. \textit{Passing context from the previous response}. In: Conversation State and Threading. Official Responses API Documentation.

\bibitem{otel-genai}
OpenTelemetry. \textit{Semantic conventions for generative AI systems}. Experimental Documentation.

\end{thebibliography}
\end{document}